\documentclass[conference, 10pt]{IEEEtran}
\usepackage{tabularx}

\ifCLASSINFOpdf
\else
\fi
\usepackage[numbers]{natbib}
\usepackage{epsfig}
\usepackage{epstopdf}
\usepackage[T1]{fontenc}
\usepackage{amssymb}
\usepackage{amsmath}
\usepackage{amsfonts}
\usepackage{verbatim}
\usepackage{graphicx}
\usepackage{hyperref}
\usepackage[usenames,dvipsnames]{xcolor}
\usepackage{algorithmic}[5]
\usepackage{algorithm}

\usepackage{array}
\usepackage{multirow}
\usepackage{caption}
\usepackage{subcaption}
\usepackage{multicol}
\usepackage{pgfplots}
\usepackage{lipsum}
\usepackage{colortbl}
\usepackage{colortbl}
\usepackage{tabularx}
\usepackage{lipsum}  
\usepackage{pifont}% http://ctan.org/pkg/pifont
\usepackage{tikz}
\newcommand*\circled[1]{\tikz[baseline=(char.base)]{
            \node[shape=circle,draw,inner sep=2pt] (char) {#1};}}

\begin{document}
\title{Power Aware Scheduling of Tasks on FPGAs in Data Centers}

\author{\IEEEauthorblockN{Rourab Paul$^1$, Marco Danelutto$^2$}
\IEEEauthorblockA{
Dept. of Computer Science, University of Pisa, Italy$^{1,2}$\\
rourab.paul@unipi.it$^1$, marco.danelutto@unipi.it$^2$} 
}

\maketitle
\vspace{-10pt}
\begin{abstract}
%Modern data centers with Field Programmable Gate Array (FPGA) Platform is becoming popular for high performance computation tasks because of its high speed, reconfiguration nature and cost effectiveness.
A variety of computing platforms like Field Programmable Gate Array (FPGA), Graphics Processing Unit (GPU) and multicore Central Processing Unit (CPU) in data centers are suitable for acceleration of data-intensive workloads. 
Especially, FPGA platforms in data centers are gaining popularity for high-performance computations due to their high speed, reconfigurable nature and cost-effectiveness. Such heterogeneous, highly parallel computational architectures in data centers, combined with high-speed communication technologies like 5G, are becoming increasingly suitable for real-time applications. However, flexibility, cost-effectiveness, high computational capabilities, and energy efficiency remain challenging issues in FPGA based data centers. 
In this context, an energy efficient scheduling solution is required to maximize the resource profitability of FPGA. 
%To fully harness the computational power offered by FPGAs, it becomes crucial to parallelize high performance algorithms and optimize their implementations. 
This paper introduces a power-aware scheduling methodology aimed at accommodating periodic hardware tasks within the available FPGAs of a data center at their potentially maximum speed. This proposed methodology guarantees the execution of these tasks using the maximum number of parallel computation units possible to implement in the FPGAs, with minimum power consumption. The proposed scheduling methodology is implemented in a data center with multiple $Alveo-50$ $Xilinx-AMD$ FPGAs and $Vitis$ $2023$ tool. The evidence from the implementation shows the proposed scheduling methodology is efficient compared to existing solutions. 
 
\end{abstract}
%\vspace{-10pt}

\IEEEpeerreviewmaketitle

\section{Introduction}
%\textcolor{red}{\%motivation}
In modern data centers, flexibility, cost-effectiveness and performance efficiency of dedicated hardware make FPGAs suitable for accelerating data-intensive applications such as video processing, financial data analysis, machine learning, and more.
The modern data centers of Microsoft \cite{microsoft}, Amazon \cite{amazon} and many other High Performance Computing Platforms (HPC) uses FPGAs in their data centers to execute different type of computationally extensive applications.
A combination of FPGA based heterogeneous platforms and high speed internet make data centers suitable for real-time applications \cite{you}. To achieve maximum profitability in FPGA-based data centers, considering factors such as power consumption, resource utilization, and meeting time budgets, employing a hardware task scheduler that can allocate a variety of tasks across multiple FPGAs can be an efficient solution.
%\textcolor{red}{\%scheduling algorithms in general purpose processor}
\par To achieve maximum profitability of reconfigurable hardware many scheduling algorithms and different solutions were proposed in last few years. Scheduling algorithms for multiprocessor platforms are designed to allocate software tasks to processors, while schedulers for reconfigurable platforms allocate hardware tasks to dedicated hardware. In this regard, the problem of scheduling hardware tasks across multiple FPGAs are quite similar with scheduling software tasks on multiprocessor platforms. Article \cite{dp-fair} demonstrates that greedy algorithms like Earliest Deadline First (EDF) \cite{edf} and Least Laxity First (LLF) \cite{llf} are unsuitable for both multiprocessor platforms and multiple FPGA scenarios.\\
%\textcolor{red}{\%possible solutions:ER-Fair}
\par In a multiprocessor platform, an Early Release Fair Scheduling (ER-fair) algorithm \cite{er-fair} and a Deadline Partitioning Fair (DP-Fair) Scheduling algorithm \cite{dp-fair} were proposed in literature to allow optimal resource consumption. Both of these algorithm takes $n_t$ tasks \{$T_1$, $T_3$, $T_3$, . . . , $T_{n_t}$ \} where each task $T_i$ has execution time $e_i$ time units, and this task needs to complete within $p_i$ time unit period. The ER-Fair \cite{er-fair} and DP-Fair \cite{dp-fair} calculate the weight $w_i$ of $T_i$ task as $\frac{e_i}{p_i}$. Both the schedulers assure the task allocations based on the weights of the tasks. If a starting time of $i^{th}$ task is $s_i$, the ER-Fair \cite{er-fair} guarantees two things for all tasks (i) For a given time slot $t$, $s_i$ $\leq$ $t$ $\leq$ $s_i+p_i$. (ii) At least  $\frac{e_i}{p_i} \times (t-s_i)$ of total $e_i$ execution time must be completed. The primary disadvantage of ER-Fair \cite{er-fair} is that the system faces overhead of uncontrolled migration and context switching which makes the system considerably expensive. In general-purpose multi-core System on Chip (SoC) platforms \cite{sjf}, context switching overhead is typically around one hundred nanoseconds, whereas in reconfigurable platforms, it ranges from ten milliseconds to one hundred milliseconds, depending on reconfiguration type, such as partial or full reconfiguration. In preemptive scheduling algorithms for reconfigurable platforms, context switching involves the complex read-back and write methodology of bitstreams through the Inter Configuration Access Port (ICAP). Therefore, uncontrolled context switching makes ER-Fair expensive on a reconfigurable platform.\\
%\textcolor{red}{\%possible solutions:DP-Fair}
DP-Fair \citep{dp-fair} aims to reduce the frequency of context switches while simultaneously maintaining scheduling optimality. DP-Fair achieves this by ensuring that ER-Fair constraints are met only at task period or deadline boundaries. DP-Fair \citep{dp-fair} partitions time slices instead of arrival time and departure time. In a given time slice, each task is assigned a workload equivalent to its proportional fair share. The Least Laxity First (LLF) \cite{llf} technique is used to schedule tasks inside a time slice. \\
%\textcolor{red}{\%problems of DP-Fair in FPGA}
\par Straight forward adoption of DP-Fair \cite{dp-fair} for periodic hardware tasks in reconfigurable hardware is not possible for its inherent architectural limitations and reconﬁguration overheads. \\
%\textcolor{red}{\% DP-Fair and DP-Wrap in FPGA}
To overcome the architectural limitations of DP-Fair, DP-Wrap integrated DP-Fair \cite{dp-fair} to minimize the maximum number of context switches to $n_f-1$ where $n_f$ is the number of reconfigurable hardware. Articles \cite{letter} and \cite{tran} approached the scheduling of periodic hard real-time task sets on both fully and partially reconfigurable systems to maximize resource utilization. Article \cite{letter} proposed scheduling strategies for two possible conditions of FPGA such as full reconfiguration and partial reconfiguration. Article \cite{letter} and \cite{tran} partitioned an FPGA into multiple tiles where lesser computational extensive tasks can be accommodated. The only difference between \cite{letter} and \cite{tran} is that \cite{tran} has inserted aperiodic tasks into the vacant time slices. Articles \cite{tran} and \cite{letter}, restricted the maximum number of context switch $nc_{max}$ by the equ. \ref{eq:ct}.
\begin{equation}
nc_{max}=\frac{n_f \times t_{slr} - \sum_{i=1}^{n_t} \frac{e_i}{p_i} \times t_{slr}}{n_f \times t_{cfg}}
\label{eq:ct}
\end{equation} 
Here $t_{slr}$ is given time slice,  $\sum_{i=1}^{n_t} \frac{e_i}{p_i} \times t_{slr}$ is the sum of shares of $n_t$ number of tasks and $t_{cfg}$ is the configuration overhead of reconfigurable hardware. The details discussion of the sum of shares is reported in Sec. \ref{sec:pre}. 
\par Schedulers in articles \cite{letter} and \cite{tran} used localized (partial) reconfiguration instead of full reconfiguration to reduce the context switching overhead. All of these articles approached the scheduling issue with light weight tasks where a single or few FPGAs are partitioned into multiple tiles to execute different tasks parallelly. On the other hand, articles \cite{microsoft}, Amazon \cite{amazon} reported a modern data center architecture connected with a stack of FPGAs which executes high processing tasks. The goals of task scheduling techniques in modern data centers differ significantly from the lightweight tasks reported in \cite{letter} and \cite{tran}. Article \cite{matteo} reported a resource aware scheduling methodology on data centers based on Heterogeneous Earliest Finish Time heuristic (HEFT-NF) algorithm. The available literature on reconfigurable data centers has not addressed the issue of energy efficiency and context switching overhead of high processing tasks.
 Article \cite{cong} reported a dataflow execution model with
Interval-Based Scheduling algorithm to balance tasks in CPUs and FPGAs of data centers. The Interval-Based Scheduling algorithm and the Earliest Finish Time heuristic are both greedy algorithms. These algorithms can potentially result in uncontrolled context switches, which cause significant costs in reconfigurable systems.

%\textcolor{red}{\% our work}
\par The primary motivation of the proposed scheduler is to maximize the profitability of FPGAs used in data centers. Given a specific time slice and a set of tasks, our proposed methodology guarantees the execution of these tasks with the maximum number of parallel computation units possible to implement in the FPGAs, with minimum power consumption.
In articles \cite{letter} and \cite{tran}, an FPGA is partitioned into numerous smaller, homogeneous tiles, with each tile running an individual task. The lightweight hardware tasks described in this literature can be accommodated using fewer slices available in these smaller tiles. On the other hand, our task scheduler handles high-processing hardware tasks that require more slices and larger tile areas. To compete current speed of CPUs and GPUs in data centers, FPGAs require many parallel computation units for current high processing tasks like data analytics algorithms, compression and decompression algorithms, etc. One of the popular $Xilinx-AMD$ FPGAs used in data centers is the Alveo50, which has $872k$ LUTs. In contrast, a task consisting of 6 parallel compression computation units and $8$ parallel decompression computation units for GZip consumes approximately one-third of the available LUTs in the $Alveo50$. 
As per article \cite{vitis}, to achieve a few GBps speed, a high processing application like GZip may need an entire $Alveo50$ FPGA. The dedicated FPGA is also required to utilize entire bandwidth of PCI for data communication. 
Therefore, this manuscript demonstrates the methodology with \textit{one tile per FPGA.} 

% Unlike lightweight tasks within different tiles of the same FPGA, high-computational tasks in data centers can be executed on dedicated FPGAs equipped with multiple parallel computation units.
 It is to be noted that the number of tiles in each FPGA can be changed depending on the resource consumption of the tasks in data centers. The effects of global reconfiguration (full reconfiguration) for \textit{one tile per FPGA} and local reconfiguration (partial reconfiguration) for \textit{multiple tiles per FPGA} in the scheduling methodology are the same. 

\par In this proposed methodology targets data centers where numerous number of FPGAs are connected with CPUs through PCI interface. Depending on the slice, LUT consumption and bandwidth of the input data channel (PCI) of a hardware task, an FPGA can accommodate a specific number of parallel computation units of a task. For a given time budget the proposed scheduler can find suitable number of parallel computation units for all given tasks depending on their power consumption and task weight. 
The contribution of this manuscript is stated below :
\begin{itemize}
\item  Given a specific time slice and a set of tasks, our proposed methodology ensures the execution of these tasks with the highest feasible number of parallel computation units within the FPGAs, achieving minimum power consumption.
\item This strategy generates multiple combinations of task sets that meet the time budget using the available hardware variants in the data center for each task. The variants of each task in all combinations of task sets have a different number of parallel computation units. Our scheduler chooses a task set combination that meets both the given time budget and has the least power consumption.
\item The multiple hardware files ($xclbin$) for different numbers of computation units of each task are pre-generated by high level synthesis tool. As per the suggestion of the proposed scheduler, the $xclbin$ files will be implemented in different FPGAs. 
\end{itemize}
The organization of the article is as follows: Sec. \ref{sec:pre} outlines our preliminaries of the problem statement. The details of our proposed scheduling algorithm and its design flow are discussed in Sec. \ref{sec:schdl}. The result and implementation of the proposed methodology is described in Sec. \ref{sec:rai}.  The conclusions are organized in Sec. \ref{sec:con}.

\section{Preliminaries}
\label{sec:pre}
Let us assume $n_t$ number of independent periodic tasks $T$=\{$T_1$, $T_2$, $T_3$, . . . , $T_{n_t}$ \} from different users arrive at data center. The data center has $n_f$ FPGAs which are connected with host CPUs through $n_f$ PCI slots. The $n_t$ tasks are to be scheduled on $n_f$ number of FPGA cards $F$=\{$F_1$, $F_3$, $F_3$, . . . , $F_{n_f}$ \}. Here $n_f$ $<<$ $n_t$. In a given time instance $t_{slr}$, to get the advantage of maximum communication bandwidth of a PCI Slot, the data center allocates one task to one FPGA. Each tasks has execution time $e_i$, completion time requirement $p_i$. The $e_i$ of task $T_i$ can be calculated by :
\begin{equation}
\sum_{i=1}^{n_t} e_i=\frac{total~data}{throughput}=\frac{td_i}{th_i}
\label{eq:th}
\end{equation}
In a data centers each task can be computed with one or multiple parallel computation units (CU). Multiple CUs can be executed parallely to achieve higher throughput. 
The execution time ($e_i$) of the same task $T_i$ varies depending on the throughput, which in turn depends on the number of computation units ($j$) allocated to it. For an example : execution times $e_1$ of task $T_1$ can be varied :
\begin{equation}
\sum_{j=1}^{n_{v1}} e_{1j}=\frac{td_1}{th_{1j}}
\label{eq:thj}
\end{equation} 
Here $n_{v1}$ is the maximum number of parallel computation unites can be accommodated in single FPGA.
Therefore, for $n_t$ tasks, equ. \ref{eq:th} can be rewritten using equ. \ref{eq:thj}.
\begin{equation}
\sum_{i=1}^{n_t} \sum_{j=1}^{n_{vi}} e_{ij}=\frac{td_i}{th_{ij}}
\label{eq:thij}
\end{equation}
Equ. \ref{eq:thij} provides multiple execution time $e_{ij}$ for different number of CUs. If there is only 1 CU for task $T_1$ with throughput $th_{11}$, the execution time $e_{11}$=$\frac{td_1}{th_{11}}$. If there are 2 CUs for task $T_1$ with throughput $th_{12}$, the execution time $e_{12}$=$\frac{td_1}{th_{12}}$. Similarly task $T_1$ with $n_{v_1}$ CUs and throughput $th_{1n_{v1}}$, the execution time $e_{1{n_v1}}$=$\frac{td_1}{th_{1n_{v1}}}$

The allocated share of task $T_i$ with $j$ number of parallel CUs and $r^{th}$ time slice $t_{slr}$ can be defined : 
\begin{equation}
\sum_{i=1}^{n_t}  shr_{ij}=\frac{e_{ij}}{p_i}  \times t_{slr} \quad \text{ : where }  1 \leq j \leq n_{vi}
\label{eq:shri}
\end{equation}

The number of active task at time slice $t_{slr}$ is the number of FPGA  $n_f$ , the sum of shares of all tasks $sum\_shr$=$\sum_{i=1}^{n_t} shr_{ij}$. The total HPC capacity is ($t_{slr}$ $\times$ $n_f$) on time slice $t_{slr}$. The primary condition of workability  to schedule the given period task will be : 
\begin{equation}
sum\_shr \leq t_{slr} \times n_f
\label{eq:con}
\end{equation}
The FPGA configuration time is $t_{cfg}$ and for the best case $n_t$ number of configurations are required to place $n_t$ number of tasks, the modified condition of workability to schedule the given tasks will be : 
\begin{equation}
sum\_shr \leq (t_{slr} \times n_f) -  (n_t \times t_{cfg})
\label{eq:conm}
\end{equation}
It is important to note that the task sets which satisfy the workability condition stated in equ. \ref{eq:conm} may not be feasible to implement in $t_{slr}$ time slices of $n_f$ FPGAs. The details discussion is reported in Sec. \ref{sec:low_power}.
The primary assumptions of this model are stated below
\begin{itemize}
%\item All tasks to be scheduled in the FPGAs of HPC are released at the same time. Each task has
%a time budget.
\item The  read and write time to or from the memory is included in the execution time $e_i$ of each task.
\item All the variants of tasks has a fixed amount of power and execution time.
\end{itemize}

\section{Proposed Scheduling Methedology}
%\section{Power Aware Deadline Partitioning Schedular for Fully Reconfigurable (PADPS-FR)}
\label{sec:schdl}
This section reports 2 parts : (A) Scheduling Algorithm and (B) Design Flow
\subsection{Scheduling Algorithm}
The proposed scheduling algorithm has three parts (1) Searching for Feasible Task Sets (2)Searching of Lowest Power Task Set (3) Placement of Lowest Power task Set in FPGA.
\begin{algorithm}[!h]
    \caption{Searching of Feasible Task Sets in PADPS-FR}
    \label{algo:fitpadps}
    \begin{algorithmic}[1]  % Add the [1] parameter to start line numbering from 1
        \REQUIRE Tasks $\{ T_1, T_2, \ldots, T_i, \ldots, T_{n_t} \}$, $t_{slr}$, $t_{cfg}$, $n_{f}$
        \ENSURE $TFS_i$
        \FOR{$i \leftarrow 1$ \TO $n_t$}
            \FOR{$j \leftarrow 1$ \TO $n_{vi}$}            
                \STATE $shr_{ij} = \frac{td_i}{th_{ij} \times p_i} \times t_{slr}$ //using equ . \ref{eq:shri}
            \ENDFOR
        \ENDFOR
        \STATE Store all combination of $shr$ and power data of $n_t$ Tasks  in $TSS$
        \FOR{$i \leftarrow 1$ \TO $(nv_1 \times nv_2 \ldots \times nv_{nt})$}
            \IF{$sum\_shr_i \leq (n_f \times t_{slr}) - (n_t \times t_{cfg})$}
                \STATE Task Fit Set: $TFS \leftarrow TSS[i]$
            \ELSE
                \STATE Task Not Fit Set: $TNFS \leftarrow TSS[i]$
            \ENDIF
        \ENDFOR
    \end{algorithmic}
\end{algorithm}
\subsubsection{Searching of Feasible Task Sets}
\label{sec:search}
This step searches for feasible task sets that can be accommodated in available time slots of $n_f$ FPGAs considering reconfiguration time $t_{cfg}$ of $n_t$ tasks. 
As stated in algorithm \ref{algo:fitpadps}, the Searching of Feasible Task Sets takes tasks $\{ T_1, T_2, \ldots, T_i, \ldots, T_{n_t} \}$, time slice $t_{slr}$, FPGA configuration time  $t_{cfg}$, and number of FPGA $n_{f}$ as inputs. As shown in the example of Table. \ref{table:exmp1}, each task $T_i$ is defined by 6 parameters : $T_i$=[ $p_i$, $td_i$, $nv_i$, $II_i$, \{ $th_{i_1}$, $th_{i_2}$, ..., $th_{nv_i}$ \}, \{ $pw_{i_1}$, $pw_{i_2}$, ..., $pw_{nv_i}$ \}]. Here $p_i$, $td_i$, $nv_i$, and $II_i$ are completion time requirement, input data size, number of variants, and initialization interval of $i^{th}$ task respectively. The  $th_{i_2}$, ..., $th_{nv_i}$ \}, \{ $pw_{i_1}$, $pw_{i_2}$, ..., $pw_{nv_i}$ \}] are throughput and power consumption of different variants of tasks.
The two for loops in lines 1-5 of algorithm \ref{algo:fitpadps} calculate the share of each variant for $n_t$ given tasks using equ. \ref{eq:shri}. Line 6 of algorithm \ref{algo:fitpadps}, calculates all possible shares along with its power consumption. It generates $nv_1 \times nv_2 \ldots \times nv_{nt}$ number of task share sets in Task Share Set list $TSS$. The $TSS$ list has $nv_1 \times nv_2 \ldots \times nv_{nt}$ rows and each row represents 1 task set. Each row contains $n_t$ shares, along with the corresponding power consumption values for $n_t$ tasks.
With the given parameters $n_t$, $n_f$, $t_{slr}$, and $t_{cfg}$, it is important to note that not all $nv_1 \times nv_2 \ldots \times nv_{nt}$ task share sets may be accommodated in $t_{slr}$ time slices of $n_f$ FPGAs based on the workability condition defined in equ. \ref{eq:conm}. Lines 7-13 of algorithm \ref{algo:fitpadps} are responsible for identifying task share sets from $TSS$ that meet the workability condition. The task share sets satisfying this condition are stored in $TFS$, while the remaining sets are stored in $TNFS$.

\begin{algorithm}[!htbp]
    \caption{Searching for Lowest Power Task Set}
    \label{algo:lowpower}
    \begin{algorithmic}[1]  % Add the [1] parameter to start line numbering from 1
        \REQUIRE $TFS$, Tasks $\{ T_1, T_2, \ldots, T_i, \ldots, T_{n_t} \}$, $t_{slr}$, $t_{cfg}$, $n_{f}$
        \ENSURE $TFS[i]$
        \STATE $Assc.~Sort~on~TFS~based~on~Total~Power~of~Tasks$
               \FOR{$i \leftarrow 1$ \TO $len(TFS)$}
       \STATE $sti=0$, $tsd=0$
        \FOR{$j \leftarrow 1$ \TO $n_f$}
        \STATE $sti$, $tsd$=find\_low\_power\_task\_set($TFS_i$, $sti$, $tsd$)
        
                  \IF {$sti ==n_t$ and $tsd==0$}
                        \RETURN $TFS[i]$
                        
                    \ENDIF
       \ENDFOR  
       \ENDFOR
         \STATE \textbf{find\_low\_power\_task\_set($sti$, $tsd$)}
          \STATE $c_j=t_{slr}$
          \FOR{$k \leftarrow sti$ \TO $n_t$}
           % \STATE $c_j=c_j-t_{cfg}-TFS[i][k]-tsd$
                    \IF {$c_j >  t_{cfg}+II_k$}
                         \IF {$c_j - t_{cfg} -TFS[i][k]<0$}
                            \STATE $sti=k$, $tsd=c_{j} -t_{cfg}$
                            \STATE \textbf{break}
                           \ELSIF{$0 \leq c_j - t_{cfg} -TFS[i][k] \leq t_{cfg}+II_k$ }
                             \STATE $sti=k+1$, $tsd=0$
                            \STATE \textbf{break}
                            \ELSE
                             \STATE $c_j=c_j-t_{cfg}-TFS[i][k]+tsd-II_k$
                         \ENDIF   
                     \ELSE
                         \STATE $sti=k$, $tsd=0$
                        \STATE \textbf{break}
                    \ENDIF                  

          \ENDFOR

         \RETURN $sti$, $tsd$     
    \end{algorithmic}
\end{algorithm}
\subsubsection{Searching for Lowest Power Task Set}
\label{sec:low_power}
While all the tasks in $TFS$ may indeed satisfy the workability condition as stated in equ. \ref{eq:conm}, it is essential to recognize that not all task sets in $TFS$ may be compatible with the provided values of $t_{slr}$, $t_{cfg}$, and $n_{f}$, and this can be attributed to two primary reasons.
\begin{itemize}
\item If an initial capacity $c_j$ of $j^{th}$ FPGA is $t_{slr}$ and share of $k^{th}$ task from $i^{th}$ task sets of $TFS$ is $TFS[i][k]$. After placement of this $k^{th}$ task, the remaining capacity of $j^{th}$ FPGA will be  : $c_j$=$c_j$-$TFS[i][k]$. Even if this new $c_j>0$, the $k+1^{th}$ task in $i^{th}$ task set of $TFS$ may not be accommodated in the same $j^{th}$ FPGA due to the reconfiguration overhead $t_{cfg}$. Therefore, the condition of task placement in $j^{th}$ FPGA will be continued until $c_j$ <= $t_{cfg}$.
\item After placement of $k^{th}$ task, if $c_j$ > $t_{cfg}$, still placing $k+1^{th}$ task in the same $j^{th}$ FPGA may not be appropriate. All the tasks have a initialization interval time $II_k$. This implies the $k^{th}$ task from $i^{th}$ task sets of $TFS$ start producing data after $t_{cfg}+II_k$ time unit. Therefore, the updated condition of task placement in $j^{th}$ FPGA will be continued until $c_j$ <= $t_{cfg}+II_k$
\end{itemize}

Because of the above-mentioned issues, there will be some cases where no task is running in a specific FPGA. This time slice is called as $NULL$ time slices as shown in Fig. \ref{fig:example1}. The \ref{eq:conm} does not consider the overhead of the NULL time slice.
Therefore, a task set with $n_t$ tasks satisfies the workability condition stated in equ. \ref{eq:conm}, may not be accommodated with $n_f$ FPGA and time slice $t_{slr}$ due to the overhead of NULL time slice. 
\par Searching for Lowest Power Task Set algorithm stated \ref{algo:lowpower} find a specific task combination from $TFS$ list which consumes the lowest power considering the issue of initialization interval.
Line 1 in algorithm \ref{algo:lowpower} sorts $TFS$ in ascending order based on the total power consumption of each task combination. Each iteration of the for loop in lines 2-10 of algorithm \ref{algo:lowpower} selects a task combination from $TFS$ and attempts to verify whether that combination is feasible within the constraints of the given time slice $t_{slr}$, the available $n_f$ FPGAs, and the initialization interval $II_k$. Each iteration of the inner for loop in lines 4-9 of algorithm \ref{algo:lowpower} calculates the amount of share from how many tasks can be accommodated within a single FPGA. Line 5 calls a function named $find\_low\_power\_task\_set()$ for $j_{th}$ FPGA which returns two parameters named as starting task index $sti$ and task share done $tsd$. The $sti$ represents the task starting index of the next $(j+1)_{th}$ FPGA. The $tsd$ represents how much share of $sti$ task is executed in current $j_{th}$ FPGA.

\par The definition of $find\_low\_power\_task\_set()$ function is reported in line 11-29 of algorithm \ref{algo:lowpower}. This function loads the capacity $c_j$ by the given time slice $t_{slr}$ for $j_{th}$ FPGA in line 12. Then it tries to fit tasks starting from $sti$ to $n_t$. The placement of tasks depends on the $c_j$. If $c_j$ is greater than  $(t_{cfg}+II_k)$, new task placement is possible, otherwise task placement is rejected and a new task must be placed in the next FPGA with task current index $sti$ (line 25). The share of the current $k_{th}$ task is not executed in the current $j_{th}$ FPGA, therefore $tsd=0$. The new task acceptance condition has three possibilities. After subtracting the task overhead $TFS[i][k]$ and its configuration time $t_{cfg}$ from capacity $c_j$ : (1) $c_j - t_{cfg} -TFS[i][k]$ < $0$,  (2) $0$ $\leq$ $c_j - t_{cfg} -TFS[i][k]$  $\leq$ $t_{cfg}+II_k$  and (3) $c_j - t_{cfg} -TFS[i][k]$ $\geq $ $t_{cfg}+II_k$.\\
\textbf{$c_j - t_{cfg} -TFS[i][k]$ < $0$ (Lines 15-17 of algorithm \ref{algo:lowpower}):}\\
This condition will be satisfied if the $k^{th}$ task in $i^{th}$ task set of $TFS$ can not be accommodated entirely in the $j^{th}$ FPGA. If the $c_j$ is the remaining capacity of $j^{th}$ FPGA, then $c_j - t_{cfg}$ amount of share of $k^{th}$ task will be executed in $j^{th}$ FPGA and remaining share of $k^{th}$ task will be executed in the next $(j+1)^{th}$ FPGA. As the task is not fully executed in $j^{th}$ FPGA, the starting tax index $tsi$ of $(j+1)^{th}$ FPGA holds the same task index $k$ and completed task $tsd$ = $c_j - t_{cfg}$. 
The satisfaction of this condition breaks the task iteration loop (lines 13-28) and proceeds to the next $(j+1)^{th}$ FPGA.\\
\textbf{$0$ $\leq$ $c_j - t_{cfg} -TFS[i][k]$  $\leq$ $t_{cfg}+II_k$ (Lines 18-20 of algorithm \ref{algo:lowpower}):}
This condition will be satisfied if the $k^{th}$ task in the $i^{th}$ task set of $TFS$ is the last task that can be fully placed in the current $j^{th}$ FPGA. In other words, the $j^{th}$ FPGA does not have enough time to configure and process data from the next $(k+1)^{th}$ task. Therefore, the starting tax index $sti$ for $(j+1)^{th}$ FPGA will next task $k+1$ and completed task share of $(k+1)^{th}$ task in $j^{th}$ FPGA : $tsd$ is $0$. \\
\textbf{$c_j - t_{cfg} -TFS[i][k]$ $\geq $ $t_{cfg}+II_k$ (Lines 21-23 of algorithm \ref{algo:lowpower}):}
This condition will be satisfied if the $k^{th}$ task in the $i^{th}$ task set of $TFS$ can be fully accommodated within the current $j^{th}$ FPGA, and this same FPGA has enough time to accommodate the next $(k+1)^{th}$ task, either fully or partially. 
\begin{algorithm}[!htbp]
    \caption{Placement of Lowest Power Task Set in FPGA}
    \label{algo:fpga_place}
    \begin{algorithmic}[1]  % Add the [1] parameter to start line numbering from 1
        \REQUIRE $TFS[i], t_{slr}$, $t_{cfg}$, $n_{f}$
        \ENSURE $FPGA~Script$
               \STATE $sti=0$, $tsd=0$
        \FOR{$j \leftarrow 1$ \TO $n_f$}
        \STATE $sti$, $tsd$=find\_low\_power\_task\_set($TFS_i$, $sti$, $tsd$)
         \STATE $fpga\_script\_j()$
       \ENDFOR  
         \STATE \textbf{find\_low\_power\_task\_set($sti$, $tsd$)}
          \STATE $c_j=t_{slr}$
          \FOR{$k \leftarrow sti$ \TO $n_t$}
           % \STATE $c_j=c_j-t_{cfg}-TFS[i][k]-tsd$
                    \IF {$c_j >  t_{cfg}+II_k$}
                         \IF {$c_j - t_{cfg} -TFS[i][k]<0$}
                            \STATE $sti=k$, $tsd=c_{j} -t_{cfg}$
                            \STATE $Split\_Task_k=TFS[i][k]$
                            \STATE $Split\_Ratio_k=tsd:TFS[i][k]-tsd$
                            \STATE $Split\_Task(Split\_Task_k,Split\_Ratio_k,data_k)$
                            \STATE \textbf{break}
                           \ELSIF{$0 \leq c_j - t_{cfg} -TFS[i][k] \leq t_{cfg}+II_k$ }
                             \STATE $sti=k+1$, $tsd=0$
                            \STATE \textbf{break}
                            \ELSE
                             \STATE $c_j=c_j-t_{cfg}-TFS[i][k]+tsd-II_k$
                         \ENDIF   
                     \ELSE
                         \STATE $sti=k$, $tsd=0$
                        \STATE \textbf{break}
                    \ENDIF                  

          \ENDFOR

         \RETURN $sti$, $tsd$      
    \end{algorithmic} 
\end{algorithm}
\subsubsection{Placement of Lowest Power Task combination in FPGA}
\label{sec:place}
Primarily algorithm \ref{algo:fpga_place} generates FPGA scripts for $n_f$ FPGAs to configure hardware files ($xclbin$), run software code (which executes on an HPC system), feed input data, and manage other application-specific configurations. This algorithm also divides the input data into the appropriate ratio for the tasks that are executed on multiple FPGAs. Algorithm \ref{algo:lowpower} takes the selected task combination by algorithm \ref{algo:lowpower}. As shown in Fig. \ref{fig:example1} task $T3$ with 2 parallel CU runs in two FPGAs $F1$ and $F2$. The $T3$ has a share : $shr$=$24$ and initilization interval : $II$=$2$. The proposed PADPS-FR algorithm executes $12$ (50\% share) ms share in FPGA $F2$ and the remaining $12$ (50\% share) ms share in FPGA $F3$. Therefore, the 24 GB input data for task $T3$ is divided into equal 1:1 parts to be fed into FPGAs $F1$ and $F3$.
Unlike algorithm \ref{algo:lowpower}, algorithm \ref{algo:fpga_place} is very similar, with the primary differences being in the data splitting activities in lines 12-14 and the FPGA script generation at line 4.

%
%\begin{figure*}[ht!]
%  \begin{tabularx}{0.8\linewidth}[t]{*{2}X}  
%    \begin{tabular}[c]{p{\linewidth}}
%\resizebox{8cm}{!}{%
%    \begin{tabular}{|c|c|c|c|c|c|c|>{\columncolor[gray]{0.8}}c|}
%        \hline
%Tasks & $p$  & $nv$ & $II$ & $td$  &  $th$&Power &shr\\
% &  (ms)  & & (ms) & (GB) &(GB/ms)&(mw)&\\\hline
%$T_1$  &60  &2   & 2  & 24& 0.5, 1   		   &\fbox{5}, 6 &\fbox{48}, \circled{24}\\\hline
%$T_2$  &60  &4   & 4  & 18& 0.5, 1, 1.5, 2     &\fbox{5}, 6, 7, 8&\fbox{36}, \circled{18},, 12, 9\\\hline
%$T_3$  &60  &4   & 2  & 48& 1, 2, 3, 4         &6, \fbox{7}, 8 , 9&48, \fbox{24}, \circled{16}, 12\\\hline
%$T_4$  &90  &4   & 4  & 36& 0.25, 0.5, 0.75, 1 &3, 4, \fbox{5}, 6 &96, 48, \fbox{32}, \circled{24}\\\hline
%$T_5$  &90  &4   & 6  & 72& 1, 2, 3, 4         &4, \fbox{4.5}, 5, 5.5 &\circled{48}, \fbox{24}, 16, 12\\\hline
%$T_6$  &90  &2   & 6  & 72& 1, 2               &4, \fbox{5} &\circled{48}, \fbox{24}\\\hline
%\multicolumn{8}{|c|}{Example 1 : $n_t$=6,  $n_f$=4,  $t_{slr}$=60 ms,  $t_{cfg}$=6 ms}\\\hline
%    \end{tabular}}
%    \label{table:exmp1}
%    \end{tabular} &
%    \centering
%    \begin{tabular}[c]{c}
%     \includegraphics[scale=0.17]{./fig/example1.eps}
%     \label{fig:example1}
%    \end{tabular} \tabularnewline
%    \captionof{table}{Example 1 : Task Set} &
%    \captionof{figure}{Example 1 : Task Scheduling of $5^{th}$ task hardware combination from $TFS$ list : [$48$, $36$, $24$, $32$, $24$, $24$]} \tabularnewline
%
%\end{tabularx}
%
%\end{figure*}
\subsection{Design Flow}
\begin{figure}[!htb]
\centering
\vspace{-10pt}
\includegraphics[scale=0.3]{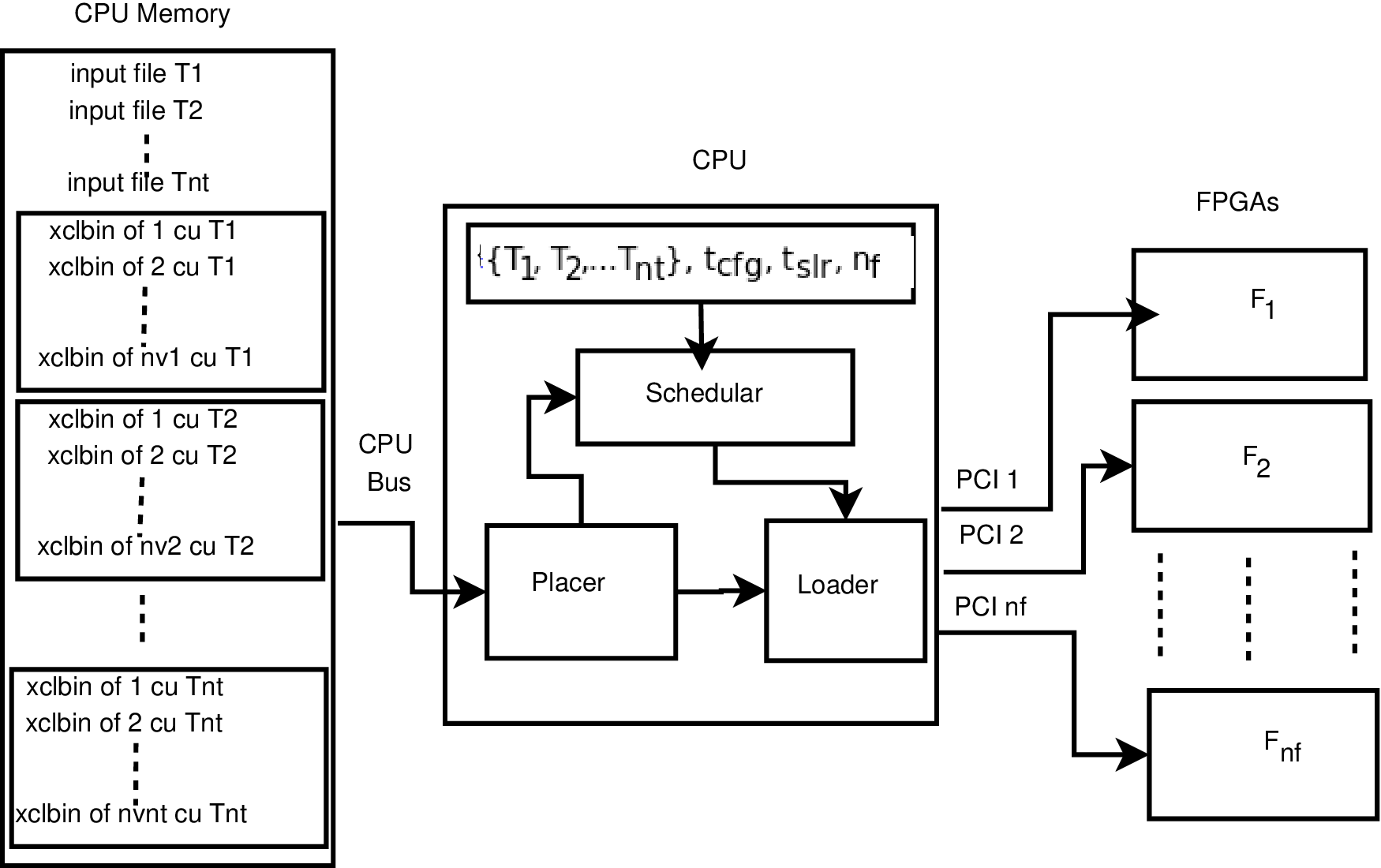}
\vspace{-5pt}
\caption{Proposed Flow}
\vspace{-10pt}
\label{fig:flow}
\end{figure}

In our proposed methodology runtime periodic hardware tasks arrive at CPU. As shown in Fig. \ref{fig:flow}, different hardware variants of these tasks are already available as $xclbin$ files in CPU memory of the data center. As described in the subsequent processes outlined in Sec. \ref{sec:search}, Sec. \ref{sec:low_power}, and Sec. \ref{sec:place}, the scheduler block selects a task combination with the lowest power consumption that meets the specified time budget using $n_f$ FPGAs.. The $placer$ block has three primary jobs (i) Fetch the $xclbin$ files of the selected task combination, (ii) Streaming input files to the appropriate $xclbin$s and FPGAs. (iii) Split of input data if preempted tasks run in multiple FPGAs. The $loader$ finally load $xclbin$ and input data to appropriate FPGAs through the PCI slots. The entire design flow is shown in Fig. \ref{fig:flow}. 

\section{Results and Implementation}
\label{sec:rai}
This section reports three subsections : A)Designs, B)Performence and C)Comparison with exitsing solutions.

\begin{figure*}[!htbp]
\centering
\vspace{-10pt}
\includegraphics[scale=0.23]{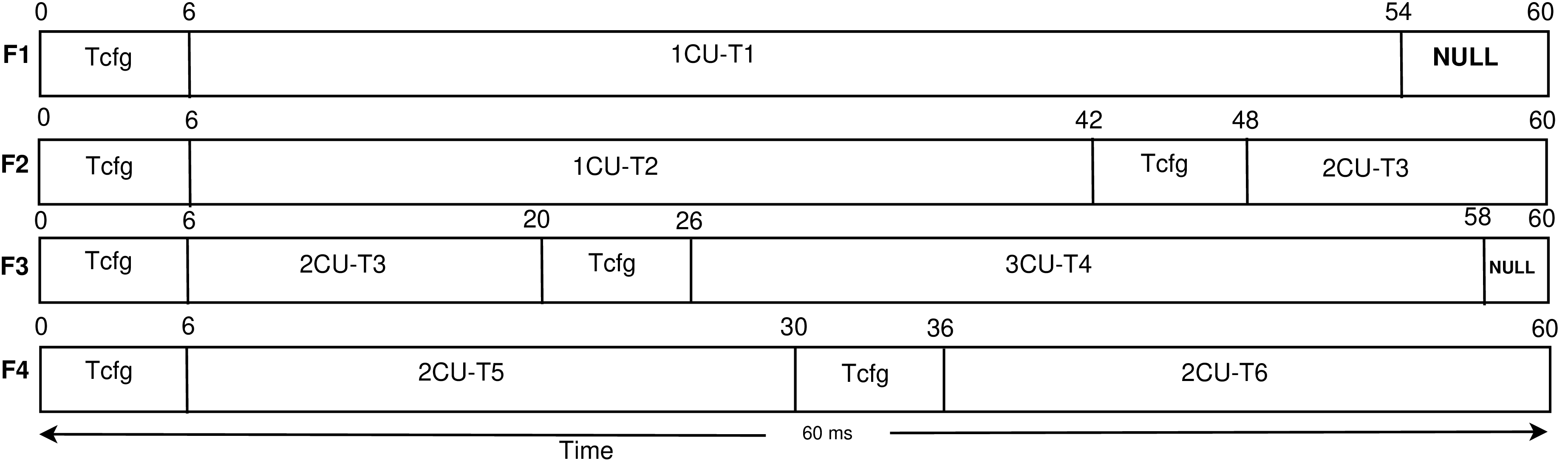}
\vspace{-5pt}
\caption{Example 1 : Task Scheduling of $5^{th}$ task hardware combination from $TFS$ list : [$48$, $36$, $24$, $32$, $24$, $24$]}
\label{fig:example1}
\end{figure*}

 \begin{table}[!htbp]
\caption{Example 1 : Task Set} % title of Table
\centering  % used for centering table
\resizebox{9cm}{!}{%
%\Rotatebox{90}{%
    \begin{tabular}{|p{1cm}|p{1cm}|p{0.5cm}|p{0.8cm}|p{1cm}|p{1.15cm}|p{1.15cm}|>{\columncolor[gray]{0.8}}p{1.5cm}|}
        \hline
Tasks & $p$  & $nv$ & $II$ & $td$  &  $th$&Power &shr\\
 &  (ms)  & & (ms) & (GB) &(GB/ms)&(mw)&\\\hline
$T_1$  &60  &2   & 2  & 24& 0.5, 1   		   &\fbox{5}, 6 &\fbox{48}, \circled{24}\\\hline
$T_2$  &60  &4   & 4  & 18& 0.5, 1, 1.5, 2     &\fbox{5}, 6, 7, 8&\fbox{36}, \circled{18},, 12, 9\\\hline
$T_3$  &60  &4   & 2  & 48& 1, 2, 3, 4         &6, \fbox{7}, 8 , 9&48, \fbox{24}, \circled{16}, 12\\\hline
$T_4$  &90  &4   & 4  & 36& 0.25, 0.5, 0.75, 1 &3, 4, \fbox{5}, 6 &96, 48, \fbox{32}, \circled{24}\\\hline
$T_5$  &90  &4   & 6  & 72& 1, 2, 3, 4         &4, \fbox{4.5}, 5, 5.5 &\circled{48}, \fbox{24}, 16, 12\\\hline
$T_6$  &90  &2   & 6  & 72& 1, 2               &4, \fbox{5} &\circled{48}, \fbox{24}\\\hline
\multicolumn{8}{|c|}{Example 1 : $n_t$=6,  $n_f$=4,  $t_{slr}$=60 ms,  $t_{cfg}$=6 ms}\\\hline
    \end{tabular}
%}
}
\label{table:exmp1} % is used to refer this table in the text
\end{table}
\subsection{Designs}
This manuscript reported 3 examples. Example 1 and 2 is simulated and Example 3 is  implemented in a data center with $2$ $Alveo-50$ FPGAs. All these 3 examples uses a hybrid environment with $Python$ $3$ and $Xilinx-AMD$ $Vitis$ $2023.1$ platform.
\subsubsection{Example 1}
\label{sec:exmp1}
Example 1 is simulated in a system with $16GB$ $RAM$, an $12th$ $Gen$ $Intel(R)$ $Core(TM)$ $i5-12400$ , and $Ubuntu$ $22.04.2$ $LTS$.
 There are $6$ hardware tasks named as $T_1$, $T_2$, $T_3$, $T_4$, $T_5$ and $T_6$. As shown in Table \ref{table:exmp1}, each hardware task has six components : period ($p$), number of variants ($nv$), initialization vector ($II$), data size to process $td$, throughputs $th$ and power consumption ($pw$). The task $T_1$ is needed to be completed with a period ($p$) of $60$ ms. The initialization interval ($II$) and the size of data ($td$) to be processed by $T_1$ are $2$ ms and $24$ GB respectively. $T_1$ has $2$ variants, where one variant has $1$ CU and other one has $2$ CUs. The throughputs of $T_1$ with $1$ CU and $2$ CU variants are $0.5$ GB/ms and $1$ GB/ms ($6^{th}$ column of table \ref{table:exmp1}) respectively. The power consumption of $T_1$ $with$ $1$ $CU$ and $T_1$ $with$ $2$ $CU$ variants are $5$ mw and $6$ mw ($7^{th}$ column of table \ref{table:exmp1}) respectively. Similarly, the throughputs and power consumptions of $T_2$, $T_3$, $T_4$, $T_5$ and $T_6$ are reported in the next five rows within the $6^{th}$ and $7^{th}$ columns of table \ref{table:exmp1}. In a given time slice boundary $t_{slr}=60$ ms, a share $shr$ of each task hardware is allocated ($8^{th}$ columns of table \ref{table:exmp1}) based on the weight of the task (line 3 algorithm \ref{algo:fitpadps}).

%\begin{figure*}[!htb]
%    \begin{minipage}{0.4\textwidth}
%        \centering
%        \resizebox{7.1cm}{!}{%
%    \begin{tabular}{|p{1cm}|p{1cm}|p{0.5cm}|p{0.8cm}|p{1cm}|p{1.15cm}|p{1.15cm}|>{\columncolor[gray]{0.8}}p{1.5cm}|}
%        \hline
%Tasks & $p$  & $nv$ & $II$ & $td$  &  $th$&Power &shr\\
% &  (ms)  & & (ms) & (GB) &(GB/ms)&(mw)&\\\hline
%$T_1$  &60  &2   & 2  & 24& 0.5, 1   		   &\fbox{5}, 6 &\fbox{48}, \circled{24}\\\hline
%$T_2$  &60  &4   & 4  & 18& 0.5, 1, 1.5, 2     &\fbox{5}, 6, 7, 8&\fbox{36}, \circled{18}, 12, 9\\\hline
%$T_3$  &60  &4   & 2  & 48& 1, 2, 3, 4         &6, \fbox{7}, 8, 9&48, \fbox{24}, \circled{16}, 12\\\hline
%$T_4$  &90  &4   & 4  & 36& 0.25, 0.5, 0.75, 1 &3, 4, \fbox{5}, 6 &96, 48, \fbox{32}, \circled{24}\\\hline
%$T_5$  &90  &4   & 6  & 72& 1, 2, 3, 4         &4, \fbox{4.5}, 5, 5.5 &\circled{48}, \fbox{24}, 16, 12\\\hline
%$T_6$  &90  &2   & 6  & 72& 1, 2               &4, \fbox{5} &\circled{48}, \fbox{24}\\\hline
%\multicolumn{8}{|c|}{Example 1 : $n_t$=6,  $n_f$=4,  $t_{slr}$=60 ms,  $t_{cfg}$=6 ms}\\\hline
%    \end{tabular}}
%       \captionof{table}{Example 1 : Task Set} 
%        \label{table:exmp1} 
%    \end{minipage}
%        \begin{minipage}{0.6\textwidth}
%        \centering
%        \includegraphics[width=\linewidth]{./fig/example1.eps} % Replace with your figure
%\caption{Example 1 : Task Scheduling of $5^{th}$ task hardware combination from $TFS$ list : [$48$, $36$, $24$, $32$, $24$, $24$]}
%\label{fig:example1}
%    \end{minipage}%
%\vspace{-15pt}    
%\end{figure*}
The number task combinations with $n_t=6$ tasks is $(nv_1 \times nv_2 \times nv_3 \times nv_4 \times nv_5 \times nv_6)$. The number of variants in our given 6 tasks are : $2$, $4$, $4$, $4$, $4$ and $2$. The number of task combinations with these $6$ tasks is $2 \times 4 \times 4 \times 4 \times 4 \times 2$ = $1024$. Therefore, the $TSS$ has $1024$ number rows. This implies the given $6$ tasks in Table \ref{table:exmp1} can be implemented in $1024$ ways. Using the workability condition stated in equ. \ref{eq:conm}, lines 7-13 of algorithm \ref{algo:fitpadps} search for task combinations that can be accommodated in 4 FPGAs.
\begin{figure*}[!htbp]
\centering
\vspace{-10pt}
\includegraphics[scale=0.23]{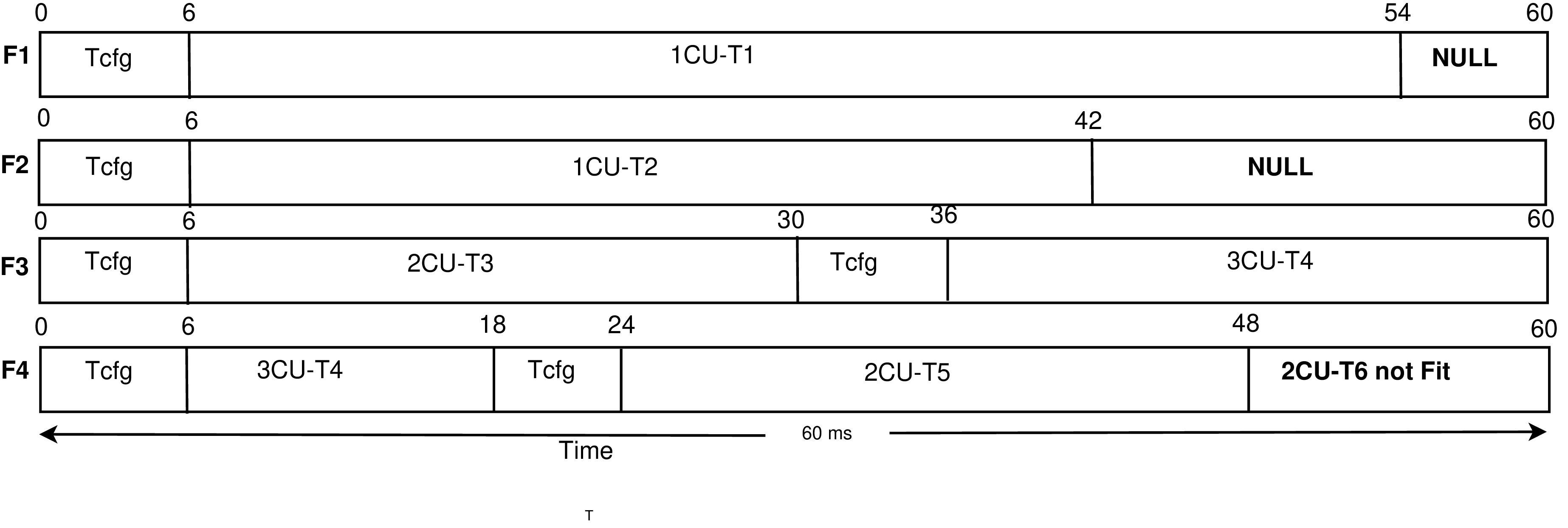}
\vspace{-5pt}
\caption{Example 2 : Task Scheduling of $5^{th}$ task hardware combination from $TFS$ list : [$48$, $36$, $24$, $32$, $24$, $24$]}
\label{fig:example2}
\end{figure*}
The lines 7-13 of algorithm \ref{algo:fitpadps} find $620$ task sets can be accommodated with given time slice $t_{slr}=60$ $ms$, number of FPGA $n_f=4$ and reconfiguration time $t_{cfg}=6$ ms. The remaining $404$ task sets violate the workability condition stated in equ. \ref{eq:conm} and cannot be accommodated with the above-mentioned parameters. Therefore, \textit{task fit set} $TFS$ and \textit{task not fit set} have $620$ and $404$ number of rows respectively. 
Let us take an arbitary task share combination from Table \ref{table:exmp1}: $[24,~18,~16,~24,~48,~48]$ (circled in $shr$ column of Table \ref{table:exmp1}). The sum of share $sum\_shr$ for this task set is $178$ which is less than  $(60 \times 4) -  (6 \times 6)=204$. This task share combination satisfies equ. \ref{eq:conm}. Therefore it will be a row among $620$ rows of $TFS$.
\par The $TFS$ list consists of $620$ rows where each row represents a unique task combination. These $620$ task combinations are then sorted in ascending order (line 1 of algorithm \ref{algo:lowpower}) based on the total power consumption of the $6$ hardware tasks. 
The algorithm \ref{algo:lowpower} further rejects $156$ task combinations from $TFS$. Therefore, the final number of rejected task combinations becomes $404+156=560$ and the total number of accepted tasks becomes $620-156=464$. The lines 2-10 in algorithm \ref{algo:lowpower} finds $5_{th}$ hardware task combination :=[$48$, $36$, $24$, $32$, $24$, $24$] (boxed in table \ref{table:exmp1})) in $TFS$ consumes least power among all $496$ tasks which can be accommodated with $n_f=4$ FPGAs and time budget $t_{slr}=60$ $ms$. 

Finally algorithm \ref{algo:fpga_place} generates $4$ scripts for $4$ FPGAs to configure FPGAs with appropriate $xclbin$ $files$, CPU scripts, input data and few other required files. As shown in Fig. \ref{fig:example1}, the $2CU-T3$ (task $T3$ with 3 parallel CUs) is executed in $2$ FPGAs. The total share of $2CU-T3$ is $24$ including $II$  $2$ $ms$. The actual data of the $2CU-T3$ task is generated for $24-2-22$ $ms$. Unlike the other $5$ tasks, $T3$ is the only preempted task that runs in multiple FPGAs, $F2$ and $F3$. As per our scheduling methodology, $12$ $ms$ share of $2CU-T3$ is executed in $F2$ FPGA. The $12$ $ms$ share of $2CU-T3$ task is divided into two parts :$II=2$ $ms$ and data generating phase $10$ $ms$.  The remaining $24-12=12$ $ms$ share of $2CU-T3$ is executed in $F3$ after $6ms$ configuration time. However, due to the reconfiguration of $2CU-T3$, the hardware again needs $2$ $ms$ $II$ followed by the data generating phase of $12$ $ms$ share. Therefore, the actual share of $2CU-T3$ in $F3$ FPGA ranges from $12$ $ms$ to $12+2=14$ $ms$. As shown in Fig. \ref{fig:flow}, the $placer$ block splits the $24$ GB input data into two separate $12$ GB files and feed these split file through the scripts of $F2$ and $F3$ generated by algorithm \ref{algo:fpga_place}.

\subsubsection{Example 2}
In an extension of example 1 stated in Sec. \ref{sec:exmp1}, if we change the $II$ of task $T3$ from $2$ $ms$ to $12$ $ms$, $2CU-T3$ cannot be placed in $F2$ FPGA. The $1CU-T2$ task is finished at $42$ $ms$. The remaining capacity of $F2$ is $60-42=18$ $ms$. The reconfiguration overhead $T_{cfg}=6$ $ms$ and $II$ of $2CU-T3$ task is $12$ $ms$. The $2CU-T3$ needs $6+12=18$ $ms$ to start producing data. Therefore placement of $2CU-T3$ in $F2$ FPGA with remaining capacity $18$ $ms$ cannot start producing data. In Fig. \ref{fig:example2}, instead of being placed in $F2$, the $2CU-T3$ will be placed in the $F3$ FPGA. Consequently, the $6$ given tasks cannot be accommodated with only $4$ FPGAs and a time budget of $60$ $ms$. As a result, the task combination [$48$, $36$, $24$, $32$, $24$, $24$] will not be selected by our proposed scheduler.

%\begin{figure*}[!htb]
%    \begin{minipage}{0.4\textwidth}
%        \centering
%        \resizebox{7.1cm}{!}{%
%    \begin{tabular}{|p{1cm}|p{1cm}|p{0.5cm}|p{0.8cm}|p{1cm}|p{1.15cm}|p{1.15cm}|>{\columncolor[gray]{0.8}}p{1.5cm}|}
%        \hline
%Tasks & $p$  & $nv$ & $II$ & $td$  &  $th$&Power &shr\\
% &  (ms)  & & (ms) & (KB) &(KB/ms)&(mw)&\\\hline
%$LZ-4$  &600  &3   & 2  & 107375 & 129.37, 165.29	198.84& 6.38, 6.55,	\fbox{6.64}
%&830, 650, \fbox{540}\\\hline
%
%$ZSTD$  &600  &2   & 2  & 107375 &244.03, 255.65 &\fbox{6.89}, 7.06 &\fbox{440}, 420\\\hline
%
%$VAdd$  &600  &4   & 2  & 19&0.12, 0.16, 0.18, 0.2&6.12, \fbox{6.21}, 6.38, 6.55 &
%159, \fbox{119},	106,	95\\\hline
%
%\multicolumn{8}{|c|}{Example 1 : $n_t$=3,  $n_f$=2,  $t_{slr}$=600 ms,  $t_{cfg}$=21ms}\\\hline
%    \end{tabular}}
%       \captionof{table}{Example 3 : Task Set} 
%        \label{table:exmp3} 
%    \end{minipage}
%        \begin{minipage}{0.6\textwidth}
%        \centering
%        \includegraphics[width=\linewidth]{./fig/example3.eps} % Replace with your figure
%\caption{Example 3 : Task Scheduling of $1^{st}$ hardware task combination from $TFS$ list : [$540$, $440$, $119$]}
%\label{fig:example3}
%    \end{minipage}%
%    \vspace{-15pt}  
%\end{figure*}
\subsubsection{Example 3}
Example 3 is implemented with $n_f=2$ $Alveo-50$ $Xilinx-AMD$ FPGAs and a HPC system configuration consisting of $128GB$ $RAM$, an $Intel(R)$ $Xeon(R)$ $CPU$ $E5-2650$ $v3$ $@$ $2.30GHz$, and $Ubuntu$ $22.04.2$ $LTS$. 
 \begin{table}[!htb]
\caption{Example 3 : Task Set} % title of Table
\centering  % used for centering table
\resizebox{9cm}{!}{%
%\Rotatebox{90}{%
    \begin{tabular}{|p{1cm}|p{1cm}|p{0.5cm}|p{0.8cm}|p{1cm}|p{1.15cm}|p{1.15cm}|>{\columncolor[gray]{0.8}}p{1.5cm}|}
        \hline
Tasks & $p$  & $nv$ & $II$ & $td$  &  $th$&Power &shr\\
 &  (ms)  & & (ms) & (KB) &(KB/ms)&(mw)&\\\hline
$LZ-4$  &600  &3   & 2  & 107375 & 129.37, 165.29	198.84& 6.38, 6.55,	\fbox{6.64}
&830, 650, \fbox{540}\\\hline

$ZSTD$  &600  &2   & 2  & 107375 &244.03, 255.65 &\fbox{6.89}, 7.06 &\fbox{440}, 420\\\hline

$VAdd$  &600  &4   & 2  & 19&0.12, 0.16, 0.18, 0.2&6.12, \fbox{6.21}, 6.38, 6.55 &
159, \fbox{119},	106,	95\\\hline

\multicolumn{8}{|c|}{Example 1 : $n_t$=3,  $n_f$=2,  $t_{slr}$=600 ms,  $t_{cfg}$=21ms}\\\hline
    \end{tabular}
%}
}
\label{table:exmp3} % is used to refer this table in the text
\end{table}
\begin{figure*}[!htb]
\centering
\includegraphics[scale=0.25]{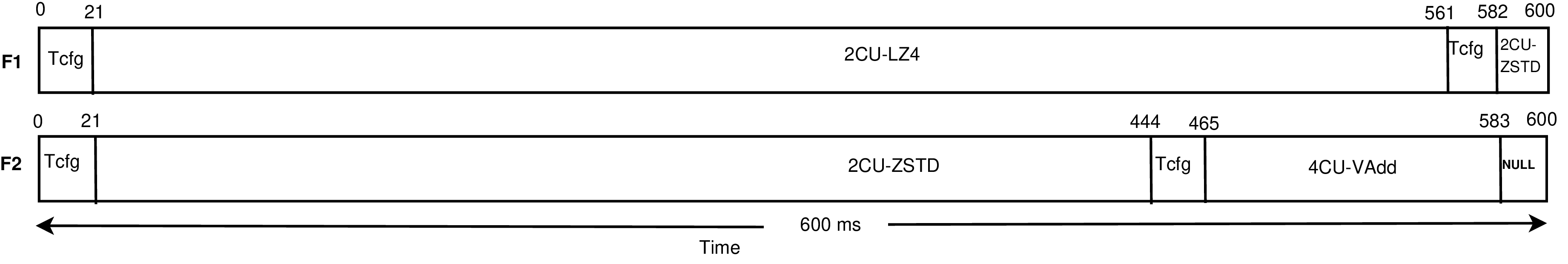}
\vspace{-5pt}
\caption{Example 3 : Task Scheduling of $1^{st}$ hardware task combination from $TFS$ list : [$540$, $440$, $119$]}
\label{fig:example3}
\end{figure*}
In example 3, there are $n_t=3$ hardware tasks: $LZ-4$, $ZSTD$, and $VAdd$. $LZ-4$ and $ZSTD$ are lossless compression algorithms, while $VAdd$ represents a vector addition process. The $LZ-4$, $ZSTD$, and $VAdd$ have 3, 2 and 4 variants respectively. The $xclbin$ files for all variants are pregenerated. The throughput and power of $LZ-4$ $with$ $1$ $CU$, $LZ-4$ $with$ $2$ $CU$ and $LZ-4$ $with$ $3$ $CU$ are mentioned in $6^{th}$ and $7^{th}$ column of Table \ref{table:exmp3} respectively. The reconfiguration time $t_{cfg}$ of $Alveo-50$ is $21$ ms. The given time slice $t_{slr}$ is 600 ms. A share $shr$ of $LZ-4$, $ZSTD$, and $VAdd$ hardware tasks are allocated ($8^{th}$ columns of table \ref{table:exmp3}) based on the weight of the tasks (line 3 algorithm \ref{algo:fitpadps}). Thereafter the line line 6 of algorithm \ref{algo:fitpadps} generates $TSS$ which has $3 \times 2 \times 4 = 24$ task combinations ($nv_{lz4}$=3, $nv_{zstd}$=2 and $nv_{vadd}$=4 ) and each task combination has $3$ tasks. Lines 7-13 of algorithm \ref{algo:fitpadps} inserts $6$ task combinations into $TFS$ which satisfies the workability condition stated in equ. \ref{eq:conm}. Rest $18$ task combinations do not satisfy the workability condition, therefore these are inserted into $TNFS$. Algorithm \ref{algo:lowpower} finds all the $6$ hardware task combinations that can be accommodated in 2 FPGAs. Therefore, the total number of rejected hardware task combinations and the total number of accepted hardware task combinations remain the same at $18$ and $6$ respectively.
Finally algorithm \ref{algo:fpga_place} generates $2$ scripts for $2$ FPGAs to configure FPGAs with appropriate xclbin files, CPU scripts, input data and a few other required files. As shown in Fig. \ref{fig:example1}, the $2CU-T3$ (task 3 with 3 parallel CUs) is executed in $2$ FPGAs. $12$ ms share of $2CU-T3$ is executed in $F2$ and rest $12$ ms share of $2CU-T3$ is executed in $F3$. As shown in Fig. \ref{fig:flow}, the $placer$ block splits the $24$ GB input data into two separate $12$ GB files and feed these split file through the scripts of $F2$ and $F3$ generated by the algorithm. 
The implementation of example 3 introduces an additional time overhead when searching for available FPGAs connected to the data center. This FPGA search function is referred to as $get\_xil\_devices()$, and it depends on the $OpenCL$ library. The timing overhead associated with this function is unpredictable, fluctuating form $\sim20$ $ms$ to $\sim90$ $ms$ for our specific case. As a result, the selected task combination [$540$, $440$, $119$] in example 3 meets the specified time budget without considering in the time overhead introduced by $get\_xil\_devices()$. The timing overhead for $get\_xil\_devices()$ may be negligible when dealing with a large volume of input data sets.

\subsection{Performance}
\begin{figure}[!htb]
\centering
\vspace{-10pt}
\includegraphics[scale=0.55]{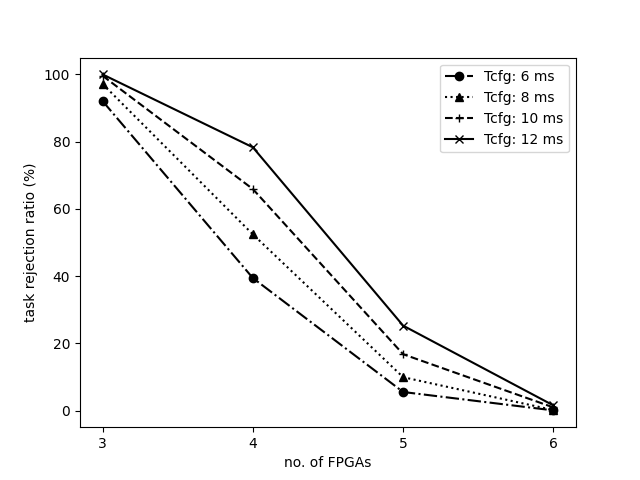}
\vspace{-5pt}
\caption{Example 1 : 1024 task set combinations : Task Rejection Ratio \% (based on equ. \ref{eq:conm}) vs No. of FPGAs}
\label{fig:task_rej}
\end{figure}

The performance of the proposed scheduler is evaluated based on three parameters (i)Task Rejection Ratio (TRR), (ii)System Work Load and (iii)Average Task Weight. The TRR is defined in equ. \ref{eq:trr}. 
\begin{equation}
TRR=\frac{no.~of~task~rejected}{total~no.~of~task~combination} \times 100 
\label{eq:trr}
\end{equation} 
The system workload is defined in equ. \ref{eq:wl}.
\begin{equation}
System~Workload=\frac{sum\_shr}{t_{slr} \times n_f} \times 100 
\label{eq:wl}
\end{equation} 
If the $system~workload$ of task combination exceeds the $system~workload~threshold$, the task combination is rejected.
The Average Task Weight is defined in equ. \ref{eq:twt}
\begin{equation}
Avg~Task~Weight=\frac {\sum_{i=1}^{n_t} \frac{e_{i}}{p_i} }{n_t}
\label{eq:twt}
\end{equation}
If the $avg~task~weight$ of task combination exceeds the $avg~task~weight~threshold$, the task combination is rejected.
In Fig. \ref{fig:task_rej}, it is observed that for a fixed reconfiguration time ($t_{cfg}$), the $TRR$ (\%) of Example 1 decreases as the number of FPGAs ($n_f$) is increased. The $TRR$ also increases with increment of $t_{cfg}$. The Task Rejection Ratio ($TRR$) approaches nearly $100\%$ when the number of FPGAs is $3$, and it drops to nearly $0\%$ when the number of FPGAs is $6$.

\begin{figure}[!htb]
\centering
\vspace{-10pt}
\includegraphics[scale=0.55]{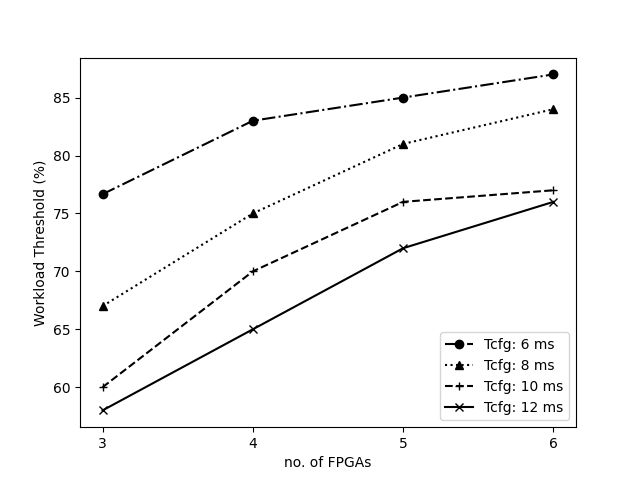}
\vspace{-5pt}
\caption{Example 1 : 1024 task set combinations : System Workload (\%) vs No. of FPGA}

\label{fig:workload}
\end{figure}
In Fig. \ref{fig:workload}, it is observed that for a fixed reconfiguration time ($t_{cfg}$), the $system~workload~threshold$ (\%) of Example 1 increases as the number of FPGAs ($n_f$) is increased. The $system~workload~threshold$ also decreases with increment of $t_{cfg}$. 
\begin{figure}[!htb]
\centering
\vspace{-10pt}
\includegraphics[scale=0.55]{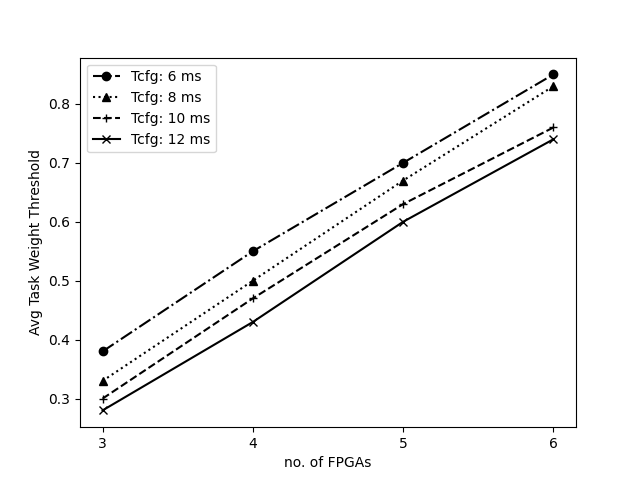}
\vspace{-5pt}
\caption{Example 1 : 1024 task set combinations : Average Task Weight  vs No. of FPGA}
\vspace{-10pt}
\label{fig:avg_task_wt}
\end{figure}
In Fig. \ref{fig:avg_task_wt}, it is observed that for a fixed reconfiguration time ($t_{cfg}$), the $avg~task~weight~threshold$ of Example 1 proportionally increases with number of FPGAs ($n_f$). The $avg~task~weight~threshold$ also decreases with increment of $t_{cfg}$.
\subsection{Comparison} 
\begin{figure}[!htb]
\centering
\vspace{-10pt}
\includegraphics[scale=0.55]{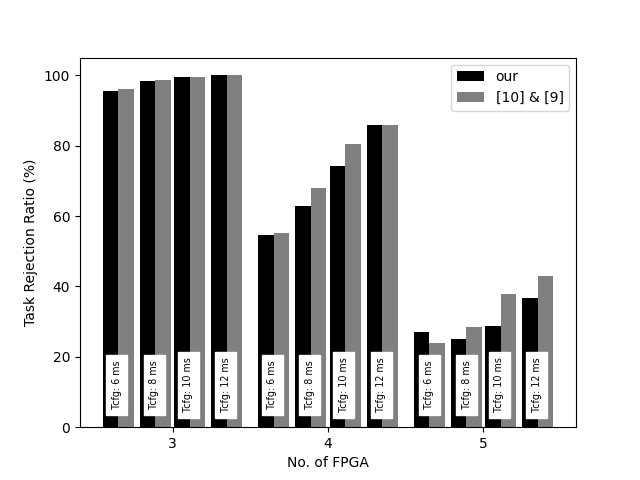}
\vspace{-5pt}
\caption{Comparison of Task Rejection Ration (\%) with Articles \cite{tran} \cite{letter}}
\vspace{-10pt}
\label{fig:com}
\end{figure}

The context switching overhead in reconfigurable hardware depends on the size of $xclbin$ $file$ \cite{hwcontext}. High processing tasks need more FPGA floor which increases the size of $xclbin$ $file$. Consequently, high processing tasks in data centers result in longer reconfiguration times, leading to higher context switching overhead. Preemptive hardware tasks involve two processes (i)context capture of current $xclbin$ files, and (ii)storing the $xclbin$ context in external memory. When these preempted hardware tasks need to be allocated back to an FPGA, there is an overhead involved in writing the previously stored $xclbin$ files into the FPGA. The reconfiguration overhead for both a fresh copy of the $xclbin$ file and a preempted task's $xclbin$ file is the same and both are referred to as $t_{cfg}$. 
For example 1, articles \cite{tran} \cite{letter} has completely ignored the context capture and context store overhead of $xclbin$ files of $2CU-T3$ task at $20^{th}$ $ms$ of $F3$ FPGA. 
Once the $2CU-T3$ is preempted in $F3$ at the $20^{th}$ $ms$, the total reconfiguration time ($t_{cfg}$) will be more than what is reported in \cite{tran} and \cite{letter} due to the context capture and context store overhead. In the case of high-processing tasks if $T4$ is $ZStd$ compression algorithm on the $Alveo-50$ FPGA, the context capture, context store, and writing of the new $xclbin$ take approximately $150$ $ms$. 
Therefore, instead of the context switching overhead $t_{cfg}$ reported in articles \cite{tran} and \cite{letter}, the actual context switching overhead for the task transition from $2CU-T3$ to $3CU-T4$ in $F3$ FPGA will be the sum of the context capture time of $2CU-T3$ $xclbin$, context store time of $2CU-T3$ $xclbin$ file, and $t_{cfg}$ of $3CU-T4$ $xclbin$ file. The captured copy of $2CU-T3$ $xclbin$ file is again downloaded in $F2$ FPGA at $42^{nd}$ $ms$. Scheduling methodology along with such expensive context switching overhead for preempted tasks in article \cite{tran} and \cite{letter} is not suitable in data centers. Our methodology does not capture and store the $xclbin$ file of half-done preempted task $2CU-T3$ at $20^{th}$ $ms$ of $F3$ FPGA. During the task transition from $1CU-T2$ to $2CU-T3$, our scheduler just writes the fresh copy of $2CU-T3$ $xclbin$ in $F2$ FPGA at $42^{nd}$ $ms$ with an extra cost of $II$. Our process methodology splits the input data files in proportion to the share allocated for $2CU-T3$ task. Fig. \ref{fig:com} shows the proposed methodology reduces the task rejection ratio compared to articles \cite{tran} \cite{letter}. The reduction in the task rejection ratio provides more flexibility in choosing a task that consumes the lowest power.

\begin{table*}[!htbp]
 \centering
	\caption{Comparison with Literature}
	%\vspace{-1.0em}    
	\label{tab:com}

\resizebox{18cm}{!}{
\begin{tabular}{|p{0.5cm}|p{1.9cm}|p{1.7cm}|p{1.5cm}|p{1.5cm}|p{2.2cm}|p{1.5cm}|p{1.2cm}|p{1.3cm}|p{6cm}|}
\hline
\textbf{Papers}& \textbf{Target Platform}   & \textbf{Implementation}  & \textbf{Scheduling Algorithm} & \textbf{Application} & \textbf{Limitations} & \textbf{Task Type} &\textbf{Energy Efficiency}& \textbf{Context Switching Restriction}& \textbf{Remarks}\\\hline   

\cite{edf}   & CPU & \center \checkmark &EDF&Low Processing Computation Platform &no hardware tasks in FPGA, unrestricted context switching  &Software& \center$\times$& \center$\times$ &mixed scheduling algorithm based on  ratemonotonic scheduling algorithm and deadline driven scheduling algorithm\\\hline 

\cite{er-fair}   & CPU & \center \checkmark &ER-Fair&Low Processing Computation Platform &no hardware tasks in FPGA, unrestricted context switching  &Software& \center$\times$& \center$\times$ &Variant of P-Fair Schedule algorithm which schedule software tasks\\\hline

\cite{letter} & FPGA-CPU, They simulated, FPGA name not mentioned & \center$\times$ &DP-Fair+DP-Wrap&Low Processing Computation Platform &Ignored context capture and context store overhead, no implementation, Energy Efficiency not considered  & Hardware& \center$\times$& \center\checkmark &Partitioned a FPGA into multiple tile. In full reconfiguration, placed multiple periodic tasks in different tiles in same time instance. In partial reconfiguration, placed multiple periodic tasks in different tiles in different time instances\\\hline 

\cite{tran} & FPGA-CPU, They simulated, FPGA name not mentioned & \center$\times$ &DP-Fair+DP-Wrap&Low Processing Computation Platform &Ignored context capture and context store overhead, no implementation, Energy Efficiency not considered  &Hardware & \center$\times$& \center\checkmark&Partitioned a FPGA into mutiple tile. In full reconfiguration, placed muiple periodic tasks in different tiles in same time instance. In partial reconfiguration, placed multiple periodic tasks in different tiles in different time instances, placed aperiodic tasks in free time slices\\\hline

\cite{matteo}    & Xilinx Spartan 7 XC7S25 FPGA-CPU & \center\checkmark &Earliest Finish Time heuristic&High Processing Computation Platform &Unrestricted Context Switching, Energy Efficiency not considered & Hardware& \center$\times$& \center$\times$  &Heuristic grouping task scheduling based on task latency\\\hline 

\cite{cong}    &  Xilinx Kintex UltraScale KU115-CPU & \center\checkmark &Interval-Based Scheduling&High Processing Computation Platform &Unrestricted Context Switching, Energy Efficiency not considered  &Software, Hardware& \center$\times$& \center$\times$  &Interval-Based Scheduling algorithm to balance tasks in CPUs and FPGAs of data centers\\\hline 

Our & Xilinx-MAD Aleveo-50 FPGA-CPU & \center\checkmark &DP-Fair+DP-Wrap&High End Computation in data center & See Conclusion at Sec. \ref{sec:con} &Hardware&\center\checkmark &\center\checkmark& See Conclusion at Sec. \ref{sec:con} \\\hline

\end{tabular}
}
\end{table*}
\section{Conclusion}
\label{sec:con}
This paper presents a scheduling methodology for high-processing hardware tasks on the reconfigurable hardware of data centers using a combination of $DP-Fair$ and $DP-Wrap$ scheduling algorithms. Given a specific time slice and a set of tasks, our proposed methodology ensures the execution of these tasks with the highest feasible number of parallel computation units within the FPGAs, while minimizing power consumption. In this paper, we have used 3 task sets named as: $Example$ $1$, $Example$ $2$ and $Example$ $3$. 
$Example$ $3$ was tested in a data center with 2 $Xilinx-AMD$ $Alveo-50$ FPGAs, while $Example$ $1$ and $Example$ $2$ were simulated. 
This work does not focus on studying scheduling possibilities for dynamic aperiodic hardware tasks. In the future, this work will also explore the scheduling of dependent periodic and aperiodic tasks. The project directory of this work is uploaded to Git Hub \cite{git}.
\bibliographystyle{unsrt}  
\bibliography{IEEEexample}

\begin{thebibliography}{10}

\bibitem{microsoft}
Adrian~M. Caulfield, Eric~S. Chung, Andrew Putnam, Hari Angepat, Jeremy Fowers,
  Michael Haselman, Stephen Heil, Matt Humphrey, Puneet Kaur, Joo-Young Kim,
  Daniel Lo, Todd Massengill, Kalin Ovtcharov, Michael Papamichael, Lisa Woods,
  Sitaram Lanka, Derek Chiou, and Doug Burger.
\newblock A cloud-scale acceleration architecture.
\newblock In {\em 2016 49th Annual IEEE/ACM International Symposium on
  Microarchitecture (MICRO)}, pages 1--13, 2016.

\bibitem{amazon}
Xinchen Liu, Wu~Liu, Huadong Ma, and Huiyuan Fu.
\newblock Large-scale vehicle re-identification in urban surveillance videos.
\newblock In {\em 2016 IEEE International Conference on Multimedia and Expo
  (ICME)}, pages 1--6, 2016.

\bibitem{you}
Feng You, Junning Qin, Keheng Zhang, Xianhui Li, Haiquan Mao, Yuxiao Zhao,
  Huayun Zhang, and Sheng Zhou.
\newblock Design and implementation of real time data center access interface
  based on big data technology.
\newblock In {\em 2017 International Conference on Computer Technology,
  Electronics and Communication (ICCTEC)}, pages 550--554, 2017.

\bibitem{dp-fair}
Greg Levin, Shelby Funk, Caitlin Sadowski, Ian Pye, and Scott Brandt.
\newblock Dp-fair: A simple model for understanding optimal multiprocessor
  scheduling.
\newblock In {\em 2010 22nd Euromicro Conference on Real-Time Systems}, pages
  3--13, 2010.

\bibitem{edf}
K.~Danne and M.~Platzner.
\newblock Periodic real-time scheduling for fpga computers.
\newblock In {\em Third International Workshop on Intelligent Solutions in
  Embedded Systems, 2005.}, pages 117--127, 2005.

\bibitem{llf}
Jinkyu Lee, Arvind Easwaran, and Insik Shin.
\newblock Llf schedulability analysis on multiprocessor platforms.
\newblock In {\em 2010 31st IEEE Real-Time Systems Symposium}, pages 25--36,
  2010.

\bibitem{er-fair}
J.H. Anderson and A.~Srinivasan.
\newblock Early-release fair scheduling.
\newblock In {\em Proceedings 12th Euromicro Conference on Real-Time Systems.
  Euromicro RTS 2000}, pages 35--43, 2000.

\bibitem{sjf}
Herbert Walder and Marco Platzner.
\newblock Online scheduling for block-partitioned reconfigurable devices.
\newblock In {\em Proceedings of the Conference on Design, Automation and Test
  in Europe - Volume 1}, DATE '03, page 10290, USA, 2003. IEEE Computer
  Society.

\bibitem{letter}
Sangeet Saha, Arnab Sarkar, and Amlan Chakrabarti.
\newblock Scheduling dynamic hard real-time task sets on fully and partially
  reconfigurable platforms.
\newblock {\em IEEE Embedded Systems Letters}, 7(1):23--26, 2015.

\bibitem{tran}
Sangeet Saha, Arnab Sarkar, Amlan Chakrabarti, and Ranjan Ghosh.
\newblock Co-scheduling persistent periodic and dynamic aperiodic real-time
  tasks on reconfigurable platforms.
\newblock {\em IEEE Transactions on Multi-Scale Computing Systems},
  4(1):41--54, 2018.

\bibitem{matteo}
Matteo Bertolino, Renaud Pacalet, Ludovic Apvrille, and Andrea Enrici.
\newblock Efficient scheduling of fpgas for cloud data center infrastructures.
\newblock In {\em 2020 23rd Euromicro Conference on Digital System Design
  (DSD)}, pages 57--64, 2020.

\bibitem{cong}
Jason Cong, Zhenman Fang, Muhuan Huang, Libo Wang, and Di~Wu.
\newblock Cpu-fpga coscheduling for big data applications.
\newblock {\em IEEE Design and Test}, 35(1):16--22, 2018.

\bibitem{vitis}
Xilinx-AMD.
\newblock Vitis data compression library.
\newblock 2022.

\bibitem{hwcontext}
Markus Happe, Andreas Traber, and Ariane Keller.
\newblock Preemptive hardware multitasking in reconos.
\newblock In Kentaro Sano, Dimitrios Soudris, Michael H{\"u}bner, and Pedro~C.
  Diniz, editors, {\em Applied Reconfigurable Computing}, pages 79--90, Cham,
  2015. Springer International Publishing.

\bibitem{git}
Rourab paul, https://github.com/rourabpaul1986/hpc\_tasks.
\newblock {\em Git Hub}, 2023.

\end{thebibliography}

\end{document}